Both exhibiting conical dispersions, graphene and Weyl Semimetals have been research frontiers for more than ten years. Recently, the research of rotating bilayer graphene has become a hot topic. However, the topological physics of two rotating Weyl structures has never been studied. In this paper, we study the topological interface states between two rotating Weyl structures based on quaternary LNOI waveguide arrays. The localization of topological interface states is experimentally investigated with both linear and nonlinear responses. Our work provides an experimental platform for studying rotating topological physics. In addition, the reported topological LNOI waveguide here can lead to potential applications in integrated nonlinear optics and quantum optics.

# Probing rotated Weyl physics on nonlinear lithium niobate-on-insulator chips

Zhi-Wei Yan,[1,‡] Qiang Wang,[2,‡] Meng Xiao,[3,*] Yu-Le Zhao[1], Shi-Ning Zhu[1], and Hui Liu[1,†]

[1]*National Laboratory of Solid State Microstructures, School of Physics, Collaborative Innovation Center of Advanced Microstructures, Nanjing University, Nanjing 210093, China.*
[2]*Division of Physics and Applied Physics, School of Physical and Mathematical Sciences, Nanyang Technological University, Singapore 637371, Singapore.*
[3]*Key Laboratory of Artificial Micro- and Nano-Structures of Ministry of Education and School of Physics and Technology, Wuhan University, Wuhan 430072, China.*

Topological photonics, featured by stable topological edge states resistant to perturbations, has been utilized to design robust integrated devices. Here, we present a study exploring the intriguing topological rotated Weyl physics in a 3D parameter space based on quaternary waveguide arrays on lithium niobate-on-insulator (LNOI) chips. Unlike previous works that focus on the Fermi arc surface states of a single Weyl structure, we can experimentally construct arbitrary interfaces between two Weyl structures whose orientations can be freely rotated in the synthetic parameter space. This intriguing system was difficult to realize in usual 3D Weyl semimetals due to lattice mismatch. We found whether the interface can host gapless topological interface states (TISs) or not, is determined by the relative rotational directions of the two Weyl structures. In the experiment, we have probed the local characteristics of the TISs through linear optical transmission and nonlinear second harmonic generation. Our study introduces a novel path to explore topological photonics on LNOI chips and various applications in integrated nonlinear and quantum optics.

*Introduction.* —Since the discovery of quantum Hall effect[1, 2], topological materials have been discovered in various physical systems, such as topological insulators[3, 4], topological semimetals[5], graphene[6], etc. The most interesting property of these topological materials is the existence of topologically protected edge states or surface states, for example, the Fermi arc surface states in Weyl semimetals. At present, the Weyl point (WP) has been observed in condensed matters[7, 8], photonic crystals[9, 10], waveguide arrays[11], and metamaterials[12-14]. In addition, with the help of synthetic dimension, people can also obtain 3D WPs with lower dimensional structures[15, 16]. Although Weyl semimetals have been widely studied, most of the researches have focused on the single Weyl crystal[7-16]. Recently, rotating bilayer graphene has become a hot topic[17]. However, the topological physics of two rotating Weyl structures has never been studied. According to our knowledge, topological interface states (TISs) between two independent Weyl structures have not been realized yet. It is because the construction of interface in condensed matters, photonic crystals, optical lattices, and metamaterials is quite challenging, which requires lattice matching and specific crystal orientation.

Branching from topological insulators, topological photonics[18-20], which hosts topological edge states[21-27], has been utilized to design robust integrated optical devices. One of the ideal platforms is the silicon-on-insulator (SOI), where various integrated topological devices have been experimentally realized[28-32], including waveguide lattices, ring resonator arrays, and photonic crystals. Besides SOI, lithium niobate-on-insulator (LNOI), as a new promising integrated-photonics platform, has attracted increasing attention due to its remarkable characteristics[33-37]. Compared with SOI, LNOI has several advantages, for instance, high second-order nonlinearity, strong electro-optic effect, and transparency in the visible region. Although some microstructures in LNOI have been reported to obtain various integrated photonic devices, including waveguides, resonant cavities and periodically poled lithium niobate (LN) structures[33-37], the territory of topological photonics on LNOI chips remains largely unexplored.

In this work, we propose an experimental study exploring synthetic rotated Weyl physics based on quaternary waveguide arrays (QWAs) on an LNOI chip. A feasible strategy is used to construct arbitrary interface between two independent Weyl structures. The interface can either host gapless TISs or not, depending on the relative rotational directions of the two Weyl structures. If they rotate in opposite directions, two types of gapless TISs arise. Otherwise, if they rotate in the same direction, only trivial interface state can exist. In experiments, we designed and fabricated spliced LNOI QWAs, and observed the TISs through linear and nonlinear optical measurements.

*Lithium niobate-on-insulator quaternary waveguide arrays*. —Figure 1(a) shows the cross-section of a LNOI QWA in a unit cell, which consists of four waveguides with the widths defined as $d_A = d_1(1+p)$, $d'_A = d_1(1-p)$, $d_B = d_2(1+q)$, and $d'_B = d_2(1-q)$, where $d_1 = 1\,\mu\text{m}$, $d_2 = 0.8\,\mu\text{m}$, and the lattice constant is $\Lambda = 2d_1 + 2d_2$. $p$ and $q$ are two independent numbers within $(-1, 1)$, which construct a 3D synthetic parameter space[38] when incorporating the 1D Bloch wave vector $k_x$.

The first four band dispersions of the fundamental transverse magnetic modes at a fixed wavelength (1040 nm) calculated by COMSOL Multiphysics (COMSOL Inc.) are plotted in Fig.1(b), where the longitudinal wave vector $k_z$ acts as the effective "energy" and $k_x$ is the 1D Bloch wave vector in unit of $k_0 = \pi/(d_1 + d_2)$. Considering the special case $p = q = 0$ as plotted with red lines, the QWA is equivalent to a binary waveguide array; thus, the two bands form a linear crossing at $k_x = 0.5k_0$ due to the band folding. The band dispersions near the crossing formed by bands 1 and 2 are enlarged in Fig. 1(c). Once deviating from $(p, q) = (0, 0)$, the degeneracy is lifted, and a band gap emerges [see the dashed black lines in Fig. 1(b)].

Considering bands 1 and 2, as most of the field distribution is located at the $A$ and $A'$ positions [see the inset of Fig. 1(c)], the $B(B')$ waveguides can be regarded as auxiliary waveguides to control the coupling strength between adjacent $A$ and $A'$ waveguides[39]. Then the QWA reduces to an effective binary optical lattice, as illustrated in the bottom of Fig. 1(a). By employing the tight-binding approximation, the Hamiltonian is

$$H = \kappa_1(q)\sum_n \left(A_n^\dagger A_n' + A_n'^\dagger A_n\right) + \kappa_2(q)\sum_n \left(A_{n+1}^\dagger A_n' + A_n'^\dagger A_{n+1}\right) + \beta_1(p)\sum_n A_n^\dagger A_n + \beta_2(p)\sum_n A_n'^\dagger A_n', \quad (1)$$

where $\kappa_{1(2)}(q)$ represent two effective coupling coefficients that are functions of $q$, and $\beta_{1(2)}(p)$ are the on-site propagation constants of the $A(A')$ waveguides that are functions of $p$. For $(p, q) = (0, 0)$, the dispersions are plotted with gray dots in Fig. 1(c), which matches well with the simulation data. By expanding near the degenerate point $(p_w, q_w, k_w) = (0, 0, 0.5k_0)$, we can get the effective Hamiltonian (Supplemental Material (SM), Sec. I[40]): $H = b\xi_p\sigma_z + 2m\xi_q v_{qx}\sigma_x - \kappa_0\xi_k\sigma_y + \beta_0\sigma_0$, where $\xi_p = p - p_w$, $\xi_q = q - q_w$, $\xi_k = (k - k_w)/k_0$ are three dimensionless coefficients, $\sigma_i (i = x, y, z)$ is the Pauli matrix, $\sigma_0$ is a $2\times 2$ identity matrix, $\beta_0 = 1.805$, $b = 0.09$, $m = 0.018$ and $\kappa_0 = 0.009$. This Hamiltonian takes a standard Weyl Hamiltonian form[41]. Figure 1(d) shows the projection of the bulk bands in the $(p, q)$ space with $k_x = 0.5k_0$. According to the definition[41], the degenerate point is actually a WP. Its 'charge' can be determined as $c = \text{sgn}(-bm\kappa_0) = -1$. In addition to this WP corresponding to $A(A')$ waveguides, we can also find another WP related to $B(B')$ waveguides if we regard $A(A')$ waveguides as auxiliary waveguides (SM, Sec. II[40]). In the following content, we only focus on the former WP. Fermi arc-like edge states related to a single WP are observed (SM, Sec. III[40]). As the parameter space is not periodic, charge neutrality is not indispensable here[16, 42]. Thus, unlike periodic systems[7-16], our system exhibits a single negative-charged WP (SM, Sec. IV[40]). This is a unique feature of synthetic dimension, which leads to incomparable advantages in controlling the surface states.

*Rotated Weyl physics.* —Here we propose a highly adjustable topological interface formed between two Weyl structures by rotating them in opposite directions. To define the rotation of them in the $k_z$ direction, we consider two rotational loops in the two parameter spaces, as illustrated in Fig. 2(a):

$$\begin{cases} \xi_{p1} = r_1 \cos\varphi_1, \\ \xi_{q1} = r_2 \sin\varphi_1, \end{cases} \begin{cases} \xi_{p2} = r_1 \cos\varphi_2, \\ \xi_{q2} = r_2 \sin\varphi_2, \end{cases} \quad (2)$$

where $\varphi_1$, $\varphi_2 \in [-\pi, \pi)$ are the rotational angles around WP1 and WP2, respectively, and $r_1$, $r_2$ are the shared radii. In such a way, we can map the original two "independent" WPs to a new $(\varphi_1, \varphi_2)$ parameter space. Numerical calculations based on the tight-binding model are carried out for the compound interfaced structure, in which $r_1 = 0.0045$, $r_2 = 0.045$, and the number of unit cells is set as 100 at each side. Figure 2(b) shows the eigenvalue spectrum in the $(\varphi_1, \varphi_2)$ space. Two colored surfaces representing two types of interface states connect the upper and lower bulk bands. We project them to the $(\varphi_1, \varphi_2)$ plane in the top inset. If the two Weyl structures rotate in opposite directions, we can define the relationship between the two angles as $\varphi_2 = -\varphi_1 + \theta$ (SM, Sec. V[40]), where $\theta$ is a constant, as illustrated by the dashed black line under periodic conditions. In this situation the topological charges of the WPs are opposite (SM, Sec. VI[40]). The loop around WP1 exhibits a counter-clockwise rotation, while the loop encircling WP2 follows a clockwise rotation, as illustrated by the arrows in Fig. 2(a). According to the bulk-edge correspondence[20], there must exist two TISs on the interface.

For further verification, we perform full-wave simulations following the path

$\varphi_2 = -\varphi_1 - 0.1\pi$ in the compound system, where $r_1 = 0.03$ and $r_2 = 0.125$, and the number of unit cells is set as 10 at each side. Figure 2(c) shows the eigenvalue spectrum as a function of $\varphi_1$, where the gray regions represent bulk bands. Two types of gapless TISs are plotted with solid red and blue dots. We choose two cases for illustration purposes [indicated by arrows in Fig. 2(c)] and plot the field intensity distributions and profiles in Fig. 2(e). For $\varphi_1 = -0.45\pi$, the interface mode has strong field confinement in the vicinity of the interface, with the majority of field intensities lying on the central $A_1'$ and $A_2'$ waveguides. For $\varphi_1 = 0.3\pi$, an exponentially localized intensity profile is revealed, with the maximum field intensity lying on the central $A_2$ waveguide. Note that we only choose $\theta = -0.1\pi$ for $\varphi_2 = -\varphi_1 + \theta$ above. Other values of $\theta$ are also analyzed in the SM, Sec. VII[40], where the phenomenon of two gapless TISs is maintained. This indicates that the interface we constructed shows great flexibility, owing to that $\xi_{p1(2)}$ and $\xi_{q1(2)}$ are synthetic momentum vectors.

Note that we only go through the curve of a fixed same size encircling each WP in the $\left(\xi_{p1(2)}, \xi_{q1(2)}\right)$ spaces. The new $\left(\varphi_1, \varphi_2, k_x\right)$ parameter space is a 3D space, where each 2D $\left(\varphi_1, k_x\right)$ slice that does not contain any Weyl nodes can be considered as a Chern insulator. The Chern number changes by two following the $\varphi_2 = -\varphi_1 + \theta$ loop, leading to two gapless TISs. If we sweep the radii of the loops, fermi arc-like interface states can be obtained. Here, the relative rotational directions determine the topological charges of Weyl nodes. On the contrary, if two Weyl nodes rotate in the same direction as $\varphi_2 = \varphi_1 + \theta'$ [see the dashed white line in Fig. 2(b)] (SM, Sec. V[40]), the topological charges are the same (SM, Sec. VIII[40]). Figure 2(d) shows an example of $\varphi_2 = \varphi_1 + 0.9\pi$. The Chern number does not change in this situation, leading to no gapless TISs, which is a trivial case. Similar phenomena take place for other values of $\theta'$.

*Experimental observation of topological interface states.* —We propose the LNOI QWAs for exploring the TISs (SM, Sec. IX[40]). Three compound QWAs were fabricated corresponding to $\varphi_1 = -0.45\pi, 0.3\pi, \pi$, where $\varphi_2 = -\varphi_1 - 0.1\pi$. The scanning electron microscopy images of one sample are presented in Figs. 3(a) and 3(b). In the experiments, we input the 1040-nm light into the QWAs and captured the output signals, which are displayed in Figs. 3(d), 3(f), and 3(h) (top panels). Their intensity profiles are displayed in bottom panels with solid black curves. For verification, numerical propagation simulations are performed in Figs. 3(c), 3(e), and 3(g). We extracted the intensity of simulations at the propagation length of experimental samples (indicated by the dashed white line) and displayed them in Figs. 3(d), 3(f), and 3(h) (bottom panels) with red bar diagrams.

For $\varphi_1 = -0.45\pi$, we take the central $A_2'$ waveguide as the input, since the maximum field intensity of TIS lies on it [see Fig. 2(e)]. The simulation shows that most of the field intensity is located near the interface. Part of the field is spreading during the propagation, due to that the input mode is not a pure TIS, but a superposition of the TIS and other bulk modes. Figure 3(d) shows good agreement between experiment and simulation, with a power maximum near the interface (dashed black line) and an exponential decay on both sides. For $\varphi_1 = 0.3\pi$, we choose the central $A_2$ waveguide as the input. The propagation simulation in Fig. 3(e) shows a singular topological defect among spreading bulk modes. In Fig. 3(f), both the experiment and simulation exhibit a power maximum near the interface, with exponentially attenuated intensities away from it, which shows a reasonable match. For $\varphi_1 = \pi$, no TIS is supported. The simulation propagation behavior in Fig. 3(g) is manifested like the well-known phenomenon of discrete diffraction[43]. The light coupled into the central $A_2'$ waveguide spreads as it propagates. The experimental measurement and the simulation follow the same diffused trends in Fig. 3(h).

The localization of light intensity near the interface indicates the existence of TISs, while the diffused light spread away from the interface shows the case without TIS (SM, Sec. X[40]). We also perform numerical calculations with random disorders introduced to the three samples. The presence of two TISs accompanied by the absence of TIS during the sampling of adiabatic pumping process is consistent with the theoretical predication of two gapless TIS bands (SM, Sec. XI[40]).

In addition to the experimental methods above, nonlinear optics can also provide a good method to probe TISs[32]. Here we use efficient second-harmonic generation (SHG) in LNOI to investigate TISs. With the same input positions as in the fundamental frequency (FF) light (1040nm) cases, the captured SHG light (520nm) at the output is shown in Fig. 4(a). We extracted the SHG intensity profiles and plotted them with curves of different colors for the three cases in Fig. 4(b). The two TIS cases ($\varphi_1 = -0.45\pi, 0.3\pi$) exhibit much stronger SHG signal than the trivial case ($\varphi_1 = \pi$), indicating a nonlinear enhancement for TISs due to the strong field localization and the immunity to fabrication imperfections. Note that SHG light exhibits a more localized distribution than FF light, which is reasonable since the coupling strength of the SHG light is weaker than that of the FF light. The SHG signals provide another effective way to observe the field localization of TISs.

*Conclusion and discussion*. —In summary, we bridge topological photonics with the LNOI platform. We demonstrate rotated Weyl physics with QWAs on an LNOI chip. We successfully construct an interface between two rotated Weyl structures whose topological relations can be flexibly tuned. The interface between two Weyl structures with opposite rotational directions supports two gapless TISs, which were experimentally observed. The local characteristics of the TISs is also probed by nonlinear SHG. In this work, we only discuss the interface state between two rotating Weyl structures in the synthetic parameter space, and this result can possibly be extended to real 3D Weyl structures. However, we have to solve the problem of lattice matching when splicing two rotated 3D Weyl lattices. It can be anticipated that they can only match for some specific rotational angles, and the mismatch for an arbitrary rotational angle will inevitably introduce significant defects. Future research on this topic is worth exploring. Beyond this, it is possible to investigate the interface states between two non-Hermitian Weyl rings if gain and loss are considered[44]. Moreover, this work introduces a new way to study linear and nonlinear topological photonics on the LNOI platform and can lead to various applications in integrated nonlinear and quantum optics.

This work was financially supported by the National Key Research and Development Program of China (Grants No. 2017YFA0205700 and No. 2017YFA0303702), the National Natural Science Foundation of China (Grants No. 11690033 and No. 11904264) and the program B for Outstanding PhD candidate of Nanjing University.

*Corresponding author.

phmxiao@whu.edu.cn

† Corresponding author.

liuhui@nju.edu.cn

‡ Z.-W.Y. and Q.W. contributed equally to this work.

process with different $\varphi_1$ s for the loop $\varphi_2 = -\varphi_1 - 0.1\pi$, which includes Refs. [42, 43].

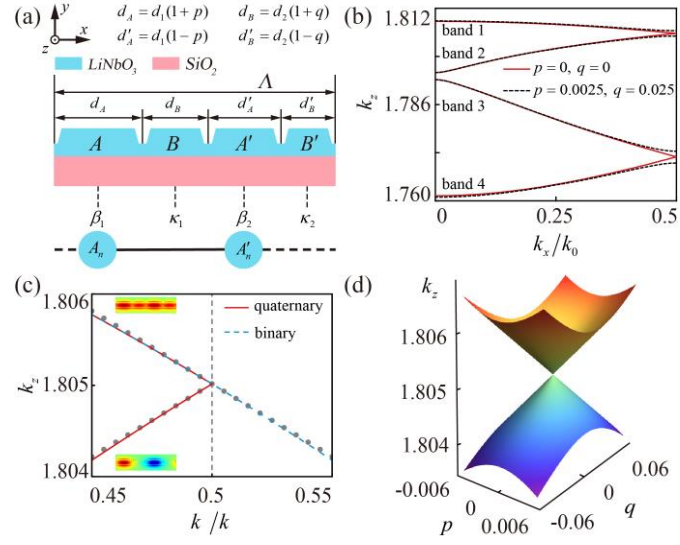

FIG. 1. Realization of a WP by the 1D QWA. (a) Schematic of a 1D QWA in a unit cell which consists of four waveguides, with 370 nm in height and a 65° sidewall angle. $p$ and $q$ are two independent numbers forming a parameter space. (b) Band dispersions at a fixed wavelength ($\lambda = 1040$ nm). Here, $k_0 = \pi/(d_1 + d_2)$ and $k_z$ is normalized by $2\pi/\lambda$. (c) The band dispersions of the QWA ($p = q = 0$) around the crossing formed by bands 1 and 2 in (b), where the solid red lines are calculated in the frame of a QWA, while the dashed blue line is from a binary waveguide array. The solid gray dots are numerical results obtained by the tight-binding approximation. (d) The band structure in the $(p, q)$ space with $k_x = 0.5k_0$ forms a conical intersection.

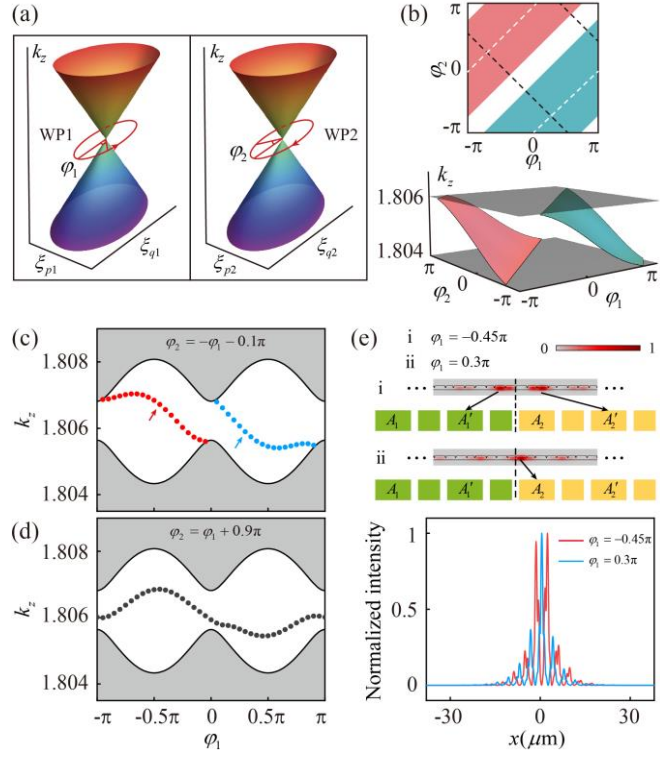

FIG. 2. TISs of the interfaced two WPs. (a) Interfacing two WPs. Two rotational loops of the same size encircling the WP are introduced in the parameter space. (b) Numerical results for small loops in (a), projected in the new $(\varphi_1, \varphi_2)$ space (bottom). Top inset: the projection to the $(\varphi_1, \varphi_2)$ plane, where the black dashed line leads to a nontrivial case while the white dashed line results in a trivial case. (c), (d) Eigenvalue spectra as a function of $\varphi_1$ by full wave simulation for large loops in (a), which follow specific paths in (b): $\varphi_2 = -\varphi_1 - 0.1\pi$ [(c), rotating in the opposite directions, nontrivial], $\varphi_2 = \varphi_1 + 0.9\pi$ [(d), rotating in the same direction, trivial). The gray regions represent bulk bands. (e) Top: field intensity distributions of the two types of TISs in (c) (indicated by the red and blue arrows). The eight central waveguides are enlarged. The black dashed line indicates the position of the interface. Bottom: the corresponding field intensity profiles.

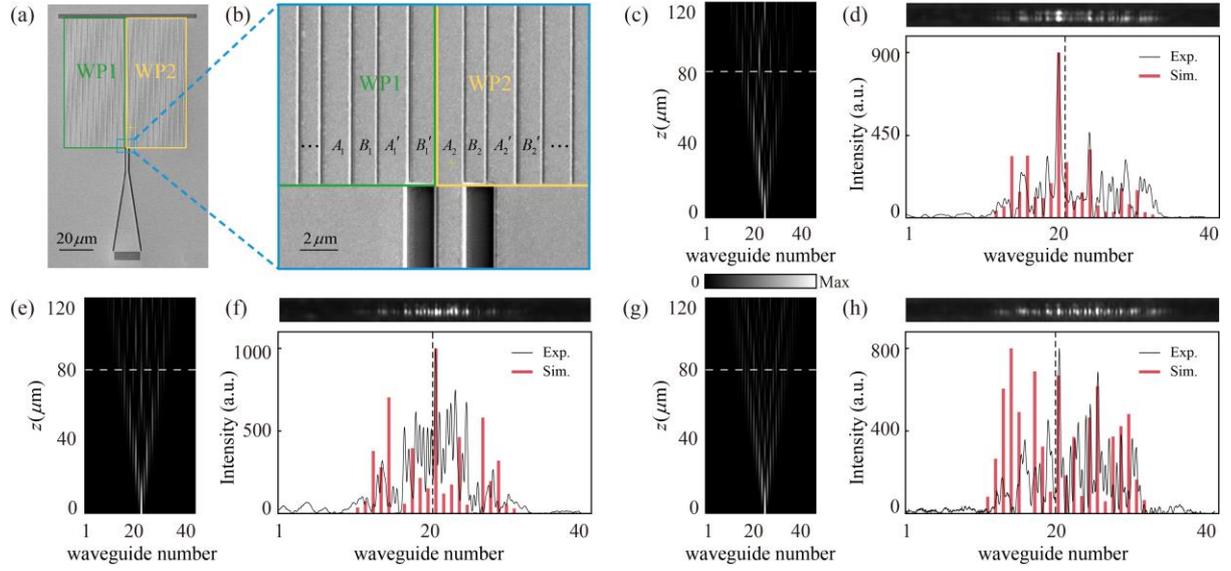

FIG. 3. Numerical and experimental results. (a) Scanning electron microscopy top view of a sample. (b) Enlarged regions in (a). (c), (e), (g) Propagation simulations. TISs cases: $\varphi_1 = -0.45\pi$ (c), $\varphi_1 = 0.3\pi$ (e); Case without TIS: $\varphi_1 = \pi$ (g), where $\varphi_2 = -\varphi_1 - 0.1\pi$. The gray color range is $[0, 0.4]$. (d), (f), (h) Experimentally detected output intensities (top) and intensity profiles (bottom) of simulation results (red bars) and experimental results (solid curves) for different cases: $\varphi_1 = -0.45\pi$ (d), $\varphi_1 = 0.3\pi$ (f), $\varphi_1 = \pi$ (h). The intensities are all normalized to the specific photoelectric conversion parameter of the sCMOS camera (Hamamatsu, ORCA-Flash 4.0, C11440-42U).

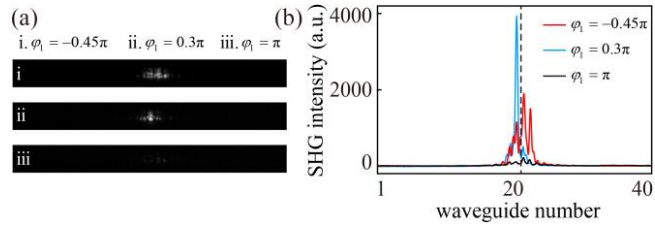

FIG. 4. Experimental second-harmonic generation (SHG) results. (a) Experimentally detected output SHG intensities for the three cases. (b) The corresponding SHG intensity profiles with colored solid curves. The intensities are all normalized to the specific photoelectric conversion parameter of the sCMOS camera (Hamamatsu, ORCA-Flash 4.0, C11440-42U).

# Supplementary Material for Probing rotated Weyl physics on nonlinear lithium niobate-on-insulator chips

Zhi-Wei Yan,[1,‡] Qiang Wang,[2,‡] Meng Xiao,[3,*] Yu-Le Zhao,[1] Shi-Ning Zhu,[1] and Hui Liu[1,†]

[1]*National Laboratory of Solid State Microstructures, School of Physics, Collaborative Innovation Center of Advanced Microstructures, Nanjing University, Nanjing 210093, China.*

[2]*Division of Physics and Applied Physics, School of Physical and Mathematical Sciences, Nanyang Technological University, Singapore 637371, Singapore.*

[3]*Key Laboratory of Artificial Micro- and Nano-Structures of Ministry of Education and School of Physics and Technology, Wuhan University, Wuhan 430072, China.*

**Section I. Effective Hamiltonian around the Weyl point (WP)**

In the main text, we have obtained the Hamiltonian by adopting the tight-binding approximation:

$$H = \kappa_1(q)\sum_n \left(A_n^\dagger A_n' + A_n'^\dagger A_n\right) + \kappa_2(q)\sum_n \left(A_{n+1}^\dagger A_n' + A_n'^\dagger A_{n+1}\right) + \beta_1(p)\sum_n A_n^\dagger A_n + \beta_2(p)\sum_n A_n'^\dagger A_n' \quad \text{(S1)}$$

where $\kappa_{1(2)}(q)$ represent two effective coupling coefficients that are functions of $q$, and $\beta_{1(2)}(p)$ are the on-site propagation constants of the $A(A')$ waveguides that are functions of $p$.

Upon Fourier transformation, we have

$$H(k) = \begin{pmatrix} \beta_1 & \kappa_1 + \kappa_2 e^{-ik} \\ \kappa_1 + \kappa_2 e^{ik} & \beta_2 \end{pmatrix}. \quad \text{(S2)}$$

Three dimensionless coefficients are defined:

$$\begin{aligned} \xi_p &= p - p_w, \\ \xi_q &= q - q_w, \\ \xi_k &= (k - k_w)/k_0. \end{aligned} \quad \text{(S3)}$$

Here, $(p_w, q_w, k_w) = (0, 0, 0.5k_0)$ denotes the position of the degenerate point.

When $\xi_p$ and $\xi_q$ are varied around the degenerate point, the widths of the waveguides change accordingly. In the vicinity of the degenerate point $(p_w, q_w, k_w) = (0, 0, 0.5k_0)$, we carry out full wave simulations using COMSOL Multiphysics to obtain $\beta_{1,2}$ and $\kappa_{1,2}$, during which the fundamental transverse magnetic (TM) modes are considered. When $\xi_p$ is varied, the widths of the $A(A')$ waveguides change and the propagation constant $\beta$ appears as an approximately linear function of $\xi_p$. We plot it in Fig. S1a, where the dots are from the full wave simulations and the red line is the fitted result. When $\xi_q$ is varied, the widths of $B(B')$ waveguides change. To obtain the effective coupling coefficients between adjacent $A$ and $A'$ waveguides, we first calculate the propagation constants of the symmetric and antisymmetric eigenmodes ($\beta_s$ and $\beta_{as}$) for a system consists of three waveguides $(ABA')$ in which $\xi_p$ is set as 0, for simplicity, near the degenerate point. Then the effective coupling coefficient can be defined as $\kappa = (\beta_s - \beta_{as})/2$. Figure. S1b shows the simulated effective coupling coefficient as a function of $\xi_q$, where the dots are from the full-wave simulations and the red line is the fitted result. So far, we obtain approximate linear relationships between $\beta_{1,2}$ and $\xi_p$, $\kappa_{1,2}$ and $\xi_q$ as

$$\begin{aligned}\beta_{1,2}(\xi_p) &= \beta_0 \pm b\xi_p, \\ \kappa_{1,2}(\xi_q) &= \kappa_0 \pm m\xi_q,\end{aligned} \quad (S4)$$

where $\beta_0 = 1.805$ and $\kappa_0 = 0.009$ are the propagation constant and the effective coupling coefficient at the exact degenerate point, respectively, $b = 0.09$ and $m = 0.018$ are slope coefficients dependent on $d_1$ and $d_2$. We adopt them to calculate the band dispersion near $k_x = 0.5k_0$ by the tight-binding model (shown in Fig.1c by the solid grey dots), which is in good agreement with the full-wave simulations.

Substituting Eq. (S4) into Eq. (S2), and expanding $H$ with respect to $(\xi_p, \xi_q, \xi_k)$ up to the first order, we finally have

$$H = (\xi_p, \xi_q, \xi_k) \mathbf{v} \begin{pmatrix} \sigma_x \\ \sigma_y \\ \sigma_z \end{pmatrix} + \beta_0 \sigma_0, \quad (S5)$$

where $\sigma_i (i = x, y, z)$ is the Pauli matrix, $\sigma_0$ is a $2 \times 2$ identity matrix, and $\mathbf{v}$ is a $3 \times 3$ real matrix

$$\mathbf{v} = \begin{pmatrix} 0 & 0 & v_{pz} \\ v_{qx} & 0 & 0 \\ 0 & v_{ky} & 0 \end{pmatrix} = \begin{pmatrix} 0 & 0 & b \\ 2m & 0 & 0 \\ 0 & -\kappa_0 & 0 \end{pmatrix}. \quad (S6)$$

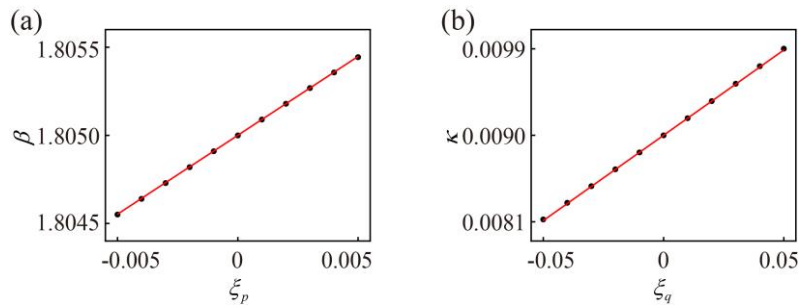

FIG. S1 (a) Propagation constant as a function of $\xi_p$. (b) Effective coupling coefficient as a function of $\xi_q$. Black dots are results from full-wave simulations, and red lines are fitted linear relationships.

## Section II. Another WP related to *B(B')* waveguides

In Fig. 1b, there is another crossing formed by bands 3 and 4. The band dispersions near this crossing are zoomed in in Fig. S2a. As $p=q=0$, we calculate this special case from both the quaternary and the binary perspectives, whose dispersions are plotted in Fig. S2a with the dashed blue line and the solid red line. The band fold thus guarantees a linear crossing along $k_x$. Insets show that most of the field distribution is located at the $B$ and $B'$ position. Instead, we regard $A(A')$ waveguides as auxiliary waveguides to control the coupling strength between adjacent $B$ and $B'$ waveguides. Results calculated by the tight-binding model are shown in Fig. S2a with solid grey dots. Figure. S2b shows the projection of the bulk bands in the $(p, q)$ space with $k_x = 0.5k_0$. Again a conical intersection appears, and the degenerate point is the WP related to $B(B')$ waveguides.

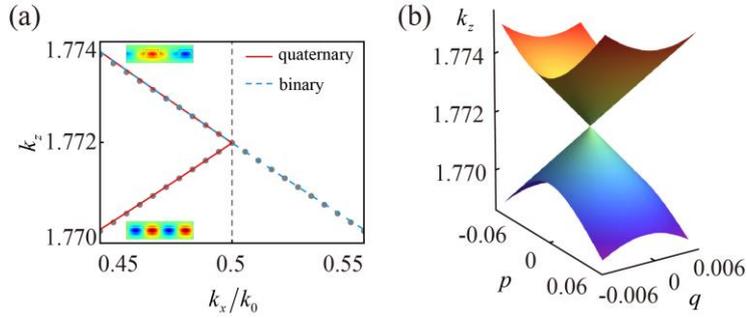

FIG. S2. The Weyl point corresponding to $B(B')$ waveguides. (a) The band dispersions of the QWA( $p=q=0$ ) around the crossing formed by band 3 and band 4 in Fig. 1(b), where the solid red lines are calculated in the frame of a QWA, while the blue dashed line is in the frame of a binary waveguide array. The solid grey dots are the corresponding numerical results obtained by the tight-binding approximation. (b) Band structure of the QWAs in the $(p, q)$ space with $k_x = 0.5k_0$. Two bands forming a conical intersection together with (a) demonstrates that the band dispersions are linear along all the three directions of the synthetic parameter space around the degenerate point, which indicates that it is also an analog of the Weyl point.

## Section III. Fermi arc-like edge states related to a single WP

We show the Fermi arc-like edge states associated with a single WP. Here, we consider an elliptical loop circling the WP in the parameter space spanned by $\xi_p$ and $\xi_q$: $\xi_p = r_1 \cos\varphi$, $\xi_q = r_2 \sin\varphi$, where $\varphi \in [-\pi, \pi)$ is the polar angle, $r_1, r_2$ control the size of the elliptical loop. An elliptic cylindrical surface forms in the parameter space, as shown in Fig. S3a. Each point on this elliptical loop corresponds to a QWA with specific parameters. We set $r_1 = 0.0045$ and $r_2 = 0.045$, and use the tight-binding model to calculate the eigenvalue spectra of different QWAs with $\varphi$ varying from $-\pi$ to $\pi$. The number of unit cells is set as 100, which is large enough. We find two crossed spirals that connect the two bulk bands, illustrated in Fig. S3a with black dashed curves. The two spirals represent a pair of topological edge states (TESs) that exist at two boundaries of the waveguide arrays respectively. As we sweep the loop's radius, the spirals at different radii are mapped to two crossed helicoids that connect the two bulk cones, which is shown in Fig. S3a. The intersection of the two helicoids forms a red line in Fig. S3a, which represents a near degeneracy of the two types of TESs. This red line can be seen as a Fermi arc emanating from the WP and ending at the boundary of the parameter space. Unlike the fact that a Fermi arc connects WPs with opposite charges in a periodic system, the edge states in our system connect a WP to the boundary of the parameter space because the total charge of the WPs does not vanish inside our parameter space.

Full-wave simulations of a larger loop with $r_1 = 0.03$ and $r_2 = 0.125$ are performed, in which the number of unit cells is set as 10. Figure S3b shows the eigenvalue spectrum as a function of the polar angle $\varphi$, in which the bulk bands are highlighted as the gray regions. There are two types of TESs in the bandgap region illustrated by solid red and blue dots. They intersect at a certain polar angle $\varphi$, which is consistent with the dashed curves for the small loop in Fig. S3a. For instance, we choose the cases $\varphi = -0.45\pi$ and $\varphi = 0.05\pi$ (indicated by arrows in Fig. S3b) and plot the field intensity distributions of the two TESs in Fig. S3c. The corresponding field intensity profiles along the *x* direction are displayed below with red and blue curves. Strong field localizations at the right (for $\varphi = -0.45\pi$) and left (for $\varphi = 0.05\pi$) boundaries are observed, which confirms the presence of TESs, where such states decay rapidly away from the boundaries.

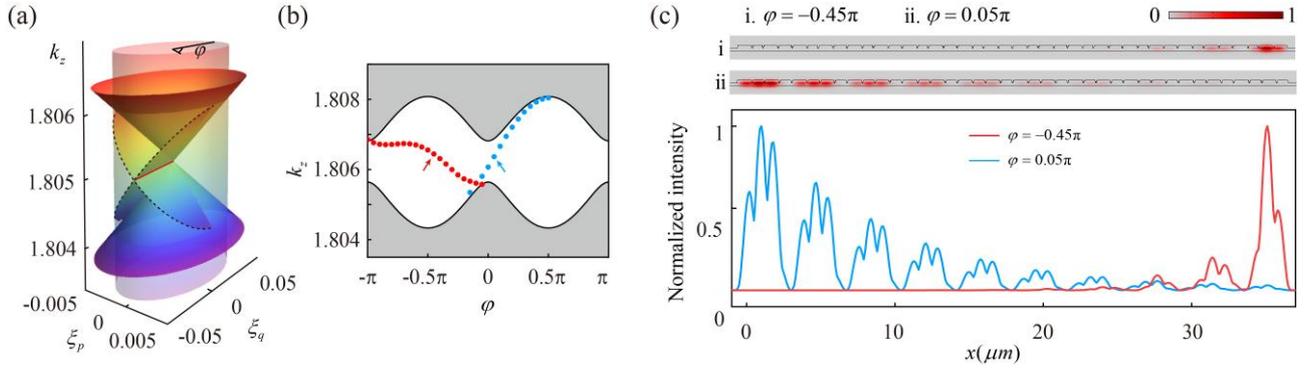

FIG. S3. Topological Edge States (TESs) of a single WP. (a) Numerical results of the TESs. Inside an elliptical loop circling the WP (shown by the elliptic cylindrical surface), two crossed helicoids that connect the two bulk cones represent two types of TESs. The intersected red line represents a Fermi arc. (b) Eigenvalue spectra as a function of the polar angle $\varphi$ by full wave simulation for a large elliptical loop circling the WP in (a). (c) Field intensity distributions of the two TESs in (b) (indicated by the red and blue arrows) and the corresponding field intensity profiles.

### Section IV. Symmetry analysis of the WP in the parameter space

For conventional WPs in the reciprocal space, the total charge of all the WPs must vanish as the reciprocal space is periodic[42], and thus WPs come in pairs with opposite charges. While the generalized parameter space $(p, q, k_x)$ has no such limits, as $p$ and $q$ were not periodic. In our system, time-reversal symmetry guarantees the Berry curvature satisfying $\mathbf{B}(p, q, k_x) = -\mathbf{B}(p, q, -k_x)$, and then WPs related by time-reversal symmetry have the same charge. Besides, our system exhibits the spatial symmetry $\varepsilon(p, q, x) = \varepsilon(-p, q, -x) = \varepsilon(p, -q, -x) = \varepsilon(-p, -q, x)$. This symmetry ensures that as long as there is a WP at $(p_w, q_w)$, there will be other WPs at $(-p_w, q_w)$, $(p_w, -q_w)$ and $(-p_w, -q_w)$, and these WPs all possess the same topological charge. In quaternary waveguide arrays (QWAs) here, we found a WP related to $A(A')$ waveguides formed by bands 1 and 2 [Fig. 1(b)] with the charge of $-1$ at $(p_w, q_w, k_w) = (0, 0, 0.5k_0)$. The above two symmetries map this WP to itself and thus the topological charge preserve, which then indicates that a single WP does exist. For another WP related to $B(B')$ waveguides formed by bands 3 and 4 at a different $k_z$ value, it can also be a single WP based on a similar analysis.

### Section V. Geometric interpretation of the rotation in the parameter space

As Fig. S4(a) shows, the quaternary waveguide arrays (QWAs) with different $p$ and $q$ parameters form the 2D parameter space. When incorporating the 1D Bloch wave vector, a Weyl point (WP) is constructed in the 3D synthetic space (Fig. S4(b)). The WP has its original chirality. We have demonstrated that it carries a charge of $\text{sgn}(-bm\kappa_0) = -1$ in the original parameter space in the main text. Without loss of generality, we can relate the charge of $-1$ to a counter-clockwise rotation as the intrinsic property of this WP.

When we form interface between systems with WP1 and WP2 following the $\varphi_2 = -\varphi_1 + \theta$ loop, an interlocked rotation emerges. There exist two kinds of external rotations as illustrated in Fig. S4(c). (i) Rotating the original $(p, q)$ space around the axis perpendicular to the $(p, q)$ plane by an arbitrary angle (upper panel), we can obtain a new WP1 in the new $(p_1, q_1)$ space whose chirality is the same as the original WP. (ii) Whereas, when we rotate the original $(p, q)$ space along the $p$ (or $q$) axis by 180° (lower panel), we obtain a new WP2 in the new $(p_2, q_2)$ space whose chirality is opposite to the original WP, as this rotation flips the sign of $q$ (or $p$) but keeps the $p$ (or $q$) and $k$ unchanged. In this way, WP1 rotates counter-clockwisely while WP2 rotates clockwisely. Figure S4(c) shows the correspondence (marked by the red point) between two parameter spaces under the mapping $\varphi_2 = -\varphi_1 + \theta$. Then, we choose the loop around WP1(2) of the same size in the $(p_{1(2)}, q_{1(2)})$ space. Interfacing two points coming from these two loops respectively forms the compound waveguide arrays, as shown by the right-most panel in Fig. S4(c). As the two WPs host opposite chiralities, the interface supports two gapless edge state when encircling the WPs.

We can analyze the situation of $\varphi_2 = \varphi_1 + \theta'$ in a similar way. Figure S4(d) shows that both the new WP1 and WP2 are constructed by rotating the original WP around the axis perpendicular to the $(p, q)$ plane. Their chiralities do not change compared with the original WP. Both WP1 and WP2 rotate counter-clockwisely after mapping. As the two WPs have the same chirality, no gapless interface states are observed when encircling them.

For conventional Weyl materials, the chirality of the Weyl node does not change if the crystal were rotated. However, here in our system, the WPs are defined in a parameter space, and the rotations are defined by going through some specific loops. To be more specific, under the condition of $\varphi_2 = -\varphi_1 + \theta$, when we rotate the original 2D $(p, q)$ space along the $p$ (or $q$) axis by 180° (lower panel of Fig. S4(c)), the sign of $q$ (or $p$) is flipped, but the $p$ (or $q$) and $k$ are kept unchanged. So after the rotation of the parameter space, the winding number of WP2 is changed, behaving like the rotational direction is changed. In comparison, such a rotation of a real 3D Weyl crystal will flip two out of the three momentum vectors, which will not change the chirality.

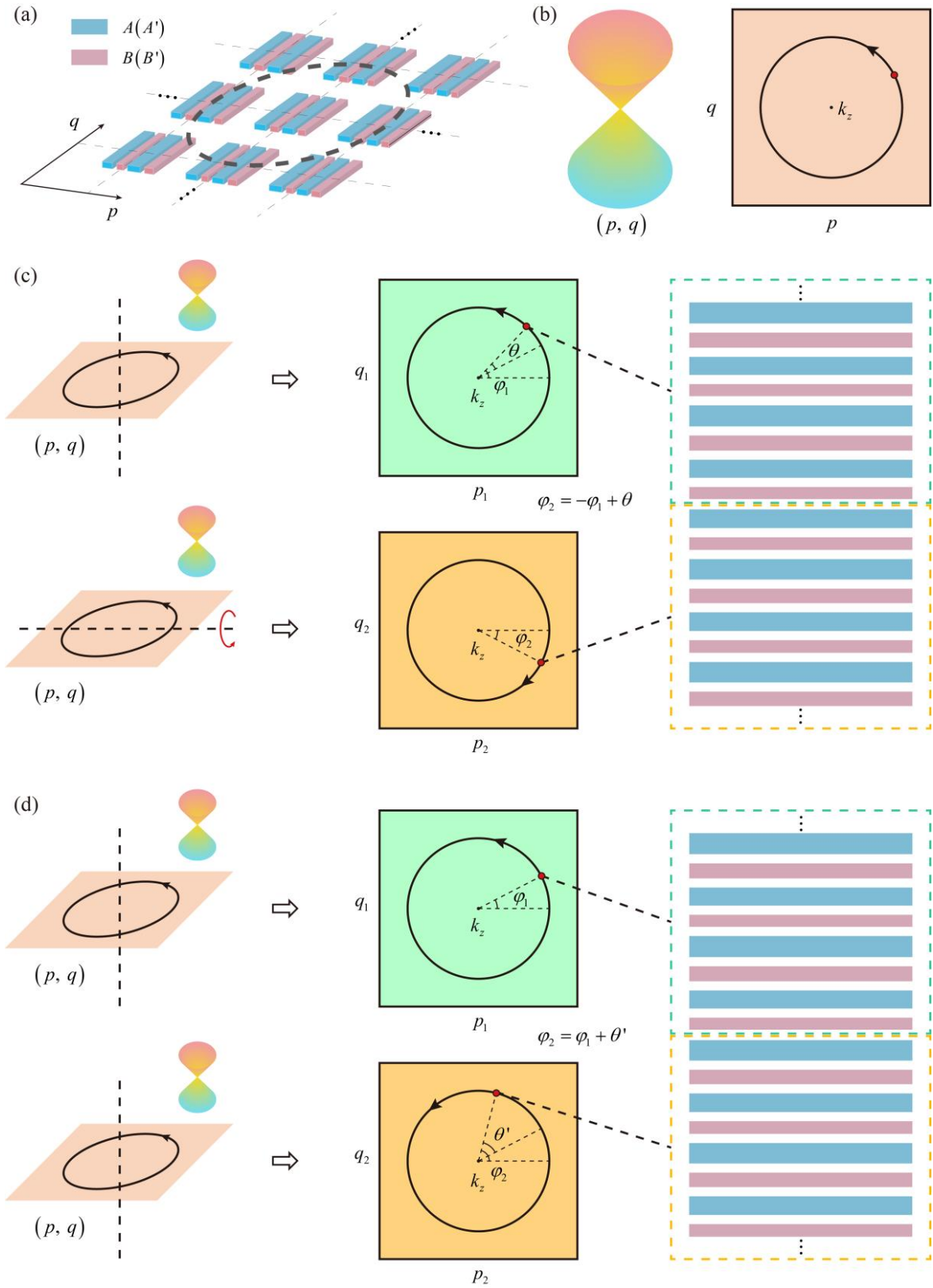

FIG. S4. (a) The quaternary waveguide arrays (QWAs) with different $p$ and $q$ parameters form a parameter space.

(b) A Weyl point (WP) is constructed in the $(p, q)$ space, whose chirality depends on a counter-clockwise rotation. (c) Following the $\varphi_2 = -\varphi_1 + \theta$ relation makes WP1 and WP2 rotate in opposite directions. (c) Following the $\varphi_2 = \varphi_1 + \theta'$ relation makes WP1 and WP2 rotate in the same direction.

**Section VI. Two WPs rotating in the opposite directions $(\varphi_2 = -\varphi_1 + \theta)$**

As discussed in the main text, the left WP rotating counter-clockwisely can be considered to carry a charge of $\mathrm{sgn}(-bm\kappa_0) = -1$ in the $(p_1, q_1)$ space. As for the right WP, opposite rotational loop based on $\varphi_2 = -\varphi_1 + \theta$ is given as

$$\begin{cases} \xi_{p2} = r_1 \cos(-\varphi_1 + \theta) = \xi_{p1} \cos\theta + \dfrac{r_1}{r_2} \xi_{q1} \sin\theta, \\ \xi_{q2} = r_2 \sin(-\varphi_1 + \theta) = \dfrac{r_2}{r_1} \xi_{p1} \sin\theta - \xi_{q1} \cos\theta. \end{cases} \quad (S7)$$

Then the Hamiltonian of the right WP in the $(p_1, q_1)$ space can be written as

$$H_2 = (\xi_{p1},\ \xi_{q1},\ \xi_{k1}) \mathbf{v}_2 \begin{pmatrix} \sigma_x \\ \sigma_y \\ \sigma_z \end{pmatrix} + \beta_0 \sigma_0, \quad (S8)$$

where $\mathbf{v}_2 = \begin{pmatrix} 2m\dfrac{r_2}{r_1}\sin\theta & 0 & b\cos\theta \\ -2m\cos\theta & 0 & b\dfrac{r_1}{r_2}\sin\theta \\ 0 & -\kappa_0 & 0 \end{pmatrix}$. So the charge of the right WP rotating clockwisely is $\mathrm{sgn}(\det[\mathbf{v}_2]) = \mathrm{sgn}(bm\kappa_0) = 1$, which is opposite of that of the left WP.

**Section VII. Eigenvalue spectrum as a function of the rotational angle $\varphi_1$ for different values of $\theta$**

We perform full wave simulations following different rotational loops $\varphi_2 = -\varphi_1 + \theta$ with $\theta = -0.9\pi,\ -0.5\pi,\ -0.1\pi,\ 0,\ 0.3\pi,\ 0.7\pi$, in which the loops' sizes are the same as the $\theta = -0.1\pi$ case in the main text. Each case shows two TIS bands in the eigenvalue spectrum in Fig. S5. As $\theta$ increases, the TIS bands exhibit a monotonic shift to the right, which can be imagined if one moves the black dashed line from bottom to top in the inset of Fig. 2b.

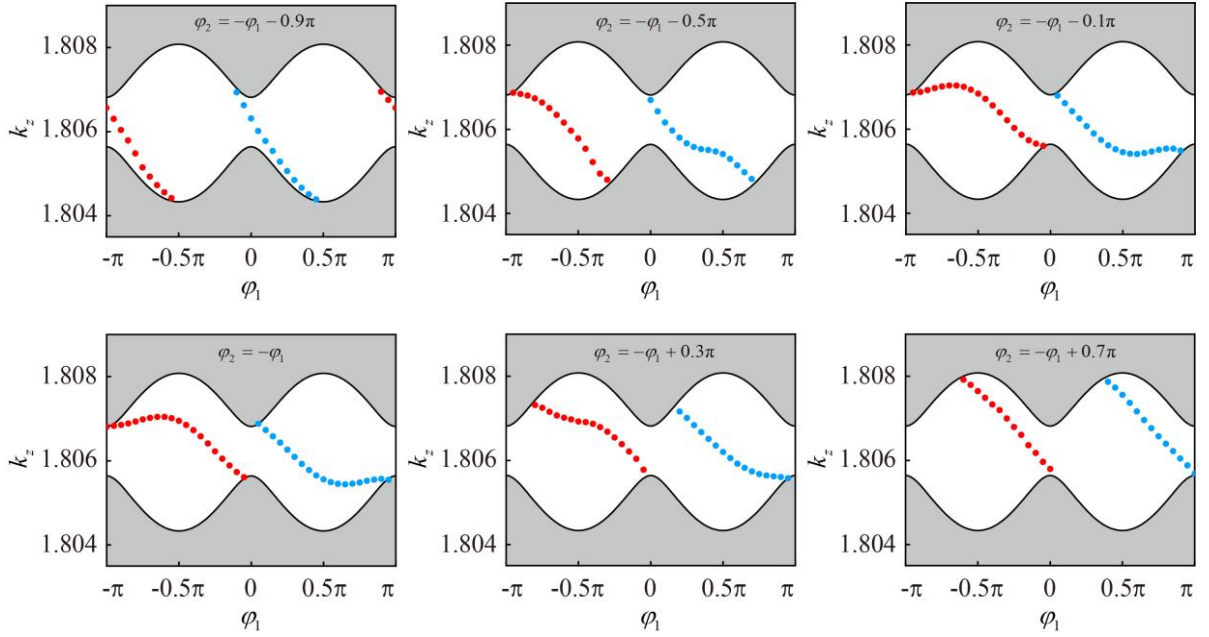

FIG. S5　Eigenvalue spectrum as a function of the rotational angle $\varphi_1$ for different values of $\theta$.

## Section VIII. Two WPs rotating in the same direction $(\varphi_2 = \varphi_1 + \theta')$

If two Weyl structures rotate in the same direction along the loop $\varphi_2 = \varphi_1 + \theta'$, the topological charges related to each side are the same. As discussed above, the left WP can be considered to carry a charge of $\text{sgn}(-bm\kappa_0) = -1$ in the $(p_1, q_1)$ space. As for the right WP, the rotational loop based on $\varphi_2 = \varphi_1 + \theta'$ is given as

$$\begin{cases} \xi_{p2} = r_1 \cos(\varphi_1 + \theta') = \xi_{p1} \cos\theta' - \dfrac{r_1}{r_2} \xi_{q1} \sin\theta', \\ \xi_{q2} = r_2 \sin(\varphi_1 + \theta') = \dfrac{r_2}{r_1} \xi_{p1} \sin\theta' + \xi_{q1} \cos\theta'. \end{cases} \quad (S9)$$

Then the Hamiltonian of the right WP can be written as

$$H_2 = \left(\xi_{p1},\ \xi_{q1},\ \xi_{k1}\right) \mathbf{v}_2 \begin{pmatrix} \sigma_x \\ \sigma_y \\ \sigma_z \end{pmatrix} + \beta_0 \sigma_0, \quad (S10)$$

where $\mathbf{v}_2 = \begin{pmatrix} 2m\dfrac{r_2}{r_1}\sin\theta' & 0 & b\cos\theta' \\ 2m\cos\theta' & 0 & -b\dfrac{r_1}{r_2}\sin\theta' \\ 0 & -\kappa_0 & 0 \end{pmatrix}$. So the charge of the right WP is

$\text{sgn}(\det[\mathbf{v}_2]) = \text{sgn}(-bm\kappa_0) = -1$, which is the same as that of the left WP.

We perform full-wave simulations following different rotational loops of $\varphi_2 = \varphi_1 + \theta'$ for $\theta' = -0.9\pi,\ -0.4\pi,\ 0,\ 0.1\pi,\ 0.6\pi,\ 0.9\pi$, in which the loops are the same as those of $\varphi_2 = -\varphi_1 + \theta$ cases. Since that no topological transition occurs at the interface, no gapless TISs that connect the upper and lower bulk bands are supported for any case here, as is depicted in Fig. S6.

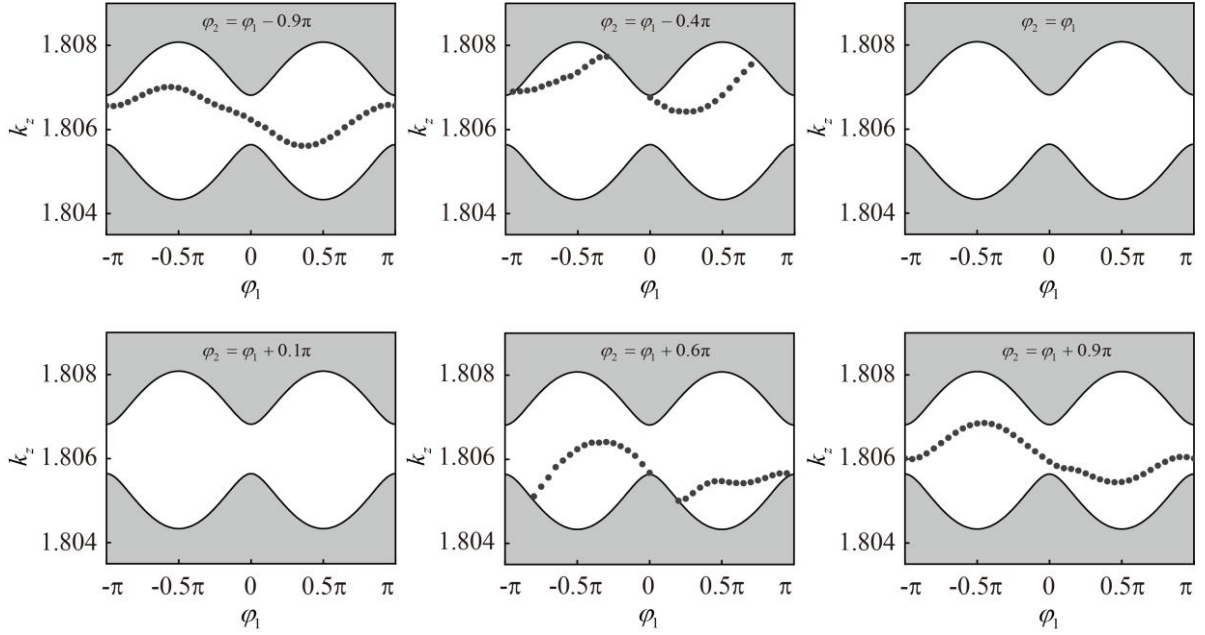

FIG. S6  Eigenvalue spectrum as a function of the rotational angle $\varphi_1$ for different values of $\theta'$.

**Section IX. Sample fabrication and measurement**

**Sample fabrication.** A commercially available z-cut LNOI wafer (NANOLN, Jinan Jingzheng Electronics Co., Ltd) was used in the fabrication. First, a 50-nm-thick silver film was sputtered onto the LNOI wafer by magnetron sputtering. Then, three waveguide arrays were drilled with a focused ion beam (FEI Dual Beam HELIOS NANOLAB 600i, 30 keV, 80 pA), together with the couple-in tapers and gratings. The lengths of the waveguide arrays are all 80 $\mu$m. The widths of the waveguides are controlled by $d_A = d_1(1+p)$, $d'_A = d_1(1-p)$, $d_B = d_2(1+q)$, $d'_B = d_2(1-q)$, where $d_1 = 1\ \mu$m, $d_2 = 0.8\ \mu$m, and the equation (4) and , in which $r_1 = 0.03$, $r_2 = 0.125$, $\varphi_2 = -\varphi_1 - 0.1\pi$. Three compound arrays were fabricated corresponding to $\varphi_1 = -0.45\pi,\ 0.3\pi,\ \pi$; for each case, the number of unit cells is set as 10 at each side of the interface. Lastly, the samples were immersed in dilute nitric acid to remove the silver film.

**Measurement.** In experiments, the laser (1040 nm) (Spectra-physics, Mai Tai HP) was focused by a microscope objective (Olympus Plan Achromat Objective 20x/0.4) and incident on the grating coupler, then coupled into the single waveguide via the taper structure. After propagating through the waveguide array, the signals were coupled out via the gratings, and collected by an sCMOS camera (Hamamatsu, ORCA-Flash 4.0, C11440-42U) through a microscope objective (Zeiss Epiplan 100x/0.75 HD microscope objective). When measuring the SHG signals, we added a $520\pm10$ nm filter before collecting the light, as shown by the dashed box in the upper panel of Fig.S7.

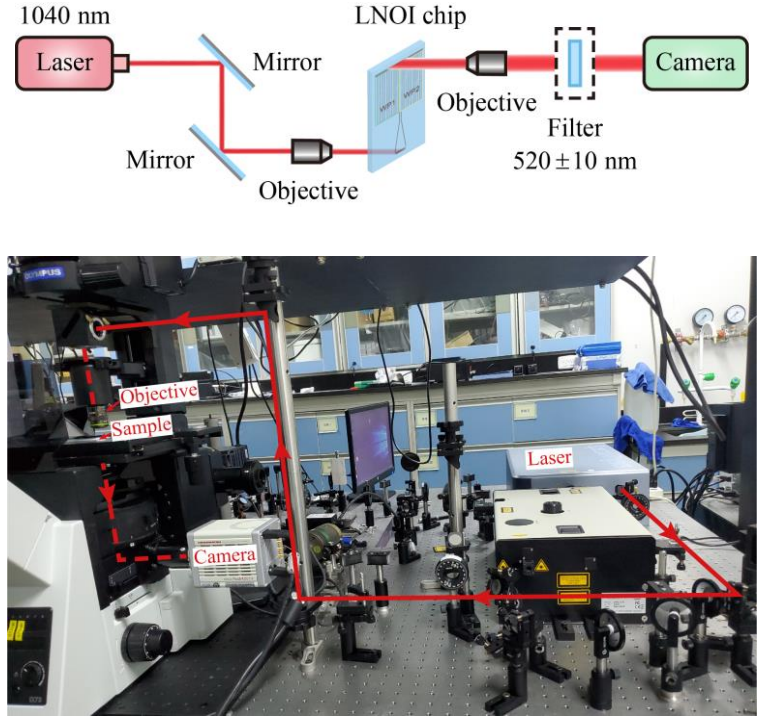

FIG. S7.  Experimental setup.

**Section X. Propagation simulations for the three samples with long distance and disorders**

The distinction between the cases with TISs and without TISs can be more obvious if we carry out propagation simulations with double propagation distance here. For the cases $\varphi_1 = -0.45\pi, 0.3\pi$ with TISs, Figs. S8 (a, b) show strong localizations near the interface despite small part of spreading bulk modes due to the excitation condition of a single waveguide input. Whereas for the case $\varphi_1 = \pi$ without TIS, the propagation behavior observed in Fig. S8(c) shows the well-known phenomenon of discrete diffraction[43], with most of the energy resides in the two sidelobes far away from the interface.

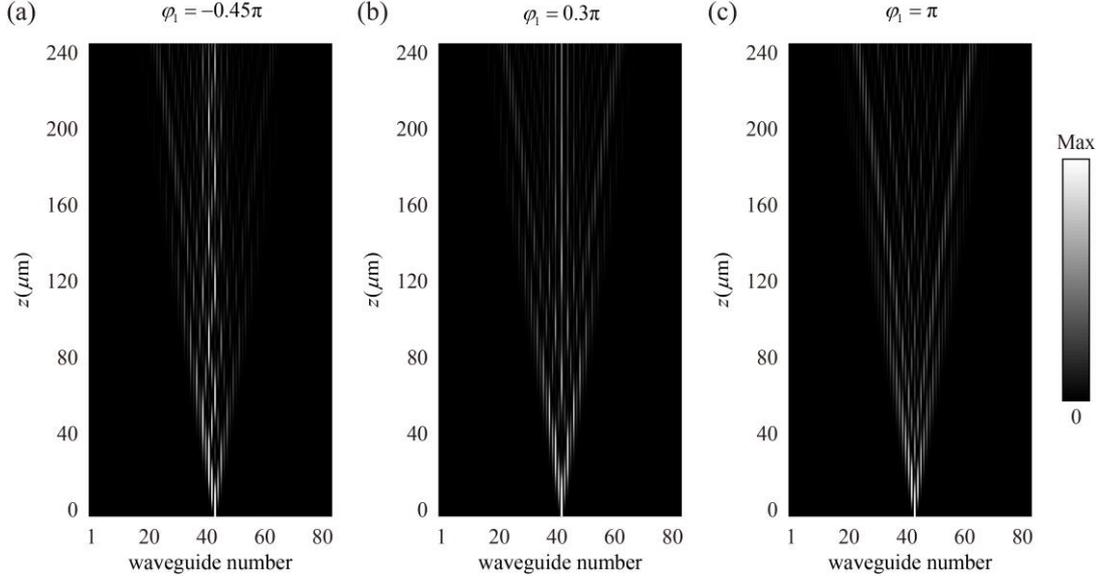

FIG. S8. Propagation simulations of the single input signal propagating through the QWAs for different cases: $\varphi_1 = -0.45\pi$ (a), $\varphi_1 = 0.3\pi$ (b), $\varphi_1 = \pi$ (c), where $\varphi_2 = -\varphi_1 - 0.1\pi$. The gray color range is $[0, 0.2]$. The waveguide numbers and propagation distance are doubled comparing with those in the main text. The position of the interface lies between the waveguide 40 and 41.

Besides, we perform numerical simulations to analyze the influence of fabrication errors on the TISs based on the experimental parameters. The widths waveguides were designed to be inside $[1000-30, 1000+30]$ nm for A-type and $[800-99, 800+99]$ nm for B-type, and the smallest distance between two adjacent waveguides was set as a constant of 30 nm. The fabrication accuracy was around $\pm 5$ nm, which was about 1/6 disorder compared with the designed parameter range. In numerical simulations, the effects of structural disorders are introduced on both the on-site propagation constants and the coupling coefficients as:

$$\begin{aligned} \beta_n' &= \beta_n + \eta \Delta\beta (\zeta_n - 0.5), \\ \kappa_n' &= \kappa_n + \eta \Delta\kappa (\zeta_n - 0.5), \end{aligned} \quad (S11)$$

where $\zeta_n$ is a uniformly distributed random number in the interval $[0, 1]$, $\beta_n$ is the original propagation constant of the waveguide $n$, $\kappa_n$ is the original coupling coefficient between waveguide $n$ and $n+1$, $\Delta\beta = 0.005217$ and $\Delta\kappa = 0.001313$ are the maximum variation range of the propagation constants and coupling coefficients, respectively, among all three samples, and $\eta$ denotes the disorder strength which is set as 1/6 here.

Figure S9 shows the average intensity distribution at the 80 $\mu$m length simulated with Eq. (1), where the error bars show the standard deviation from the mean value of 30 samples. Obviously, the results for all three samples are stable. For the TIS cases $\varphi_1 = -0.45\pi, 0.3\pi$, the characteristics of the localization near the interface preserve; For the case $\varphi_1 = \pi$ without TIS, the intensity distribution is still diffused. Based on the analysis above, the TISs manifest good robustness against fabrication errors in the compound LNOI waveguide arrays, which preserve good localization near the interface despite random disorders. We thus expect that the main features in Figs. 3(d,f,h) in the main text are reproducible on other arrays with the same designed parameters.

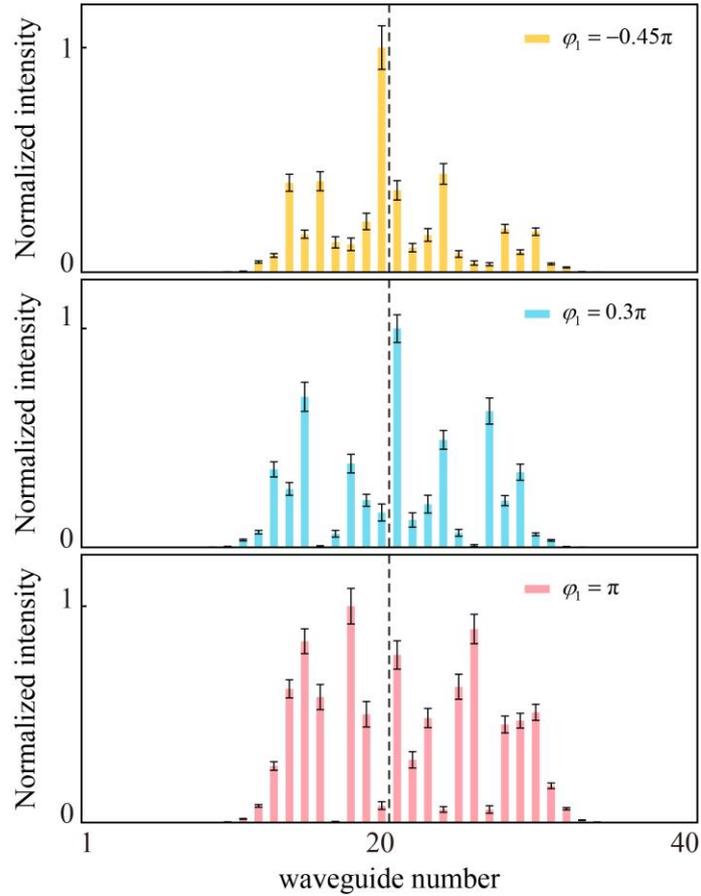

FIG. S9. Simulated average intensity distribution in the presence of disorder at 80 $\mu$m for the three cases considered in the main text, i.e., $\varphi_1 = -0.45\pi$, $\varphi_1 = 0.3\pi$, $\varphi_1 = \pi$, where $\varphi_2 = -\varphi_1 - 0.1\pi$ is fixed. The error bars show the standard deviation from the mean value of 30 samples. The positions of the interface are indicated by the black dashed line.

**Section XI. Propagation simulations showing the adiabatic process with different $\varphi_1$s for the loop $\varphi_2 = -\varphi_1 - 0.1\pi$.**

We perform several propagation simulations to show the adiabatic process with different $\varphi_1$s. Under the condition of $\varphi_2 = -\varphi_1 - 0.1\pi$, we choose ten $\varphi_1$s with the parameters considered in Fig. 3(c), as indicated by the dashed lines in Fig. S10, with four points in either TIS band. Figure S11 show the results on the condition of single input. When $\varphi_1$ increases from $-0.8\pi$ to $-0.2\pi$, the propagation patterns undergo an evolution process that the localization strength of TIS near the interface changes from weak to strong and then to weak again. This phenomenon indicates the possible gapless connectivity to bulk bands of this TIS band, which is consistent with the spectra in Fig. S10. This is because that when $\varphi_1$ increases from $-0.8\pi$ to $-0.2\pi$, the TIS in the bandgap is close to the upper bulk band at first, and become far away from the bulk band, and then approach the lower bulk band, leading to the change of the localization strength of TISs. Similar phenomena occur when $\varphi_1$ increases from $0.2\pi$ to $0.8\pi$, as shown by the propagation patterns in the bottom panels of Fig.S11. The cases C5 ($\varphi_1 = 0$) and C10 ($\varphi_1 = \pi$) that do not support TIS show spreading propagation away from the interface. The adiabatic process with these $\varphi_1$s indicates the existence of two TIS bands.

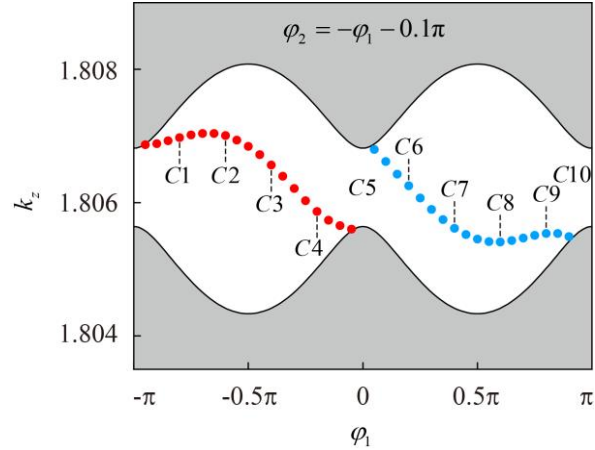

FIG. S10. Eigenvalue spectra as a function of $\varphi_1$ by full wave simulation following the path $\varphi_2 = -\varphi_1 - 0.1\pi$. We choose eight cases support TISs (C1-4, C6-9) indicated by the dashed lines and two cases without TIS (C5, C10) to perform propagation simulations.

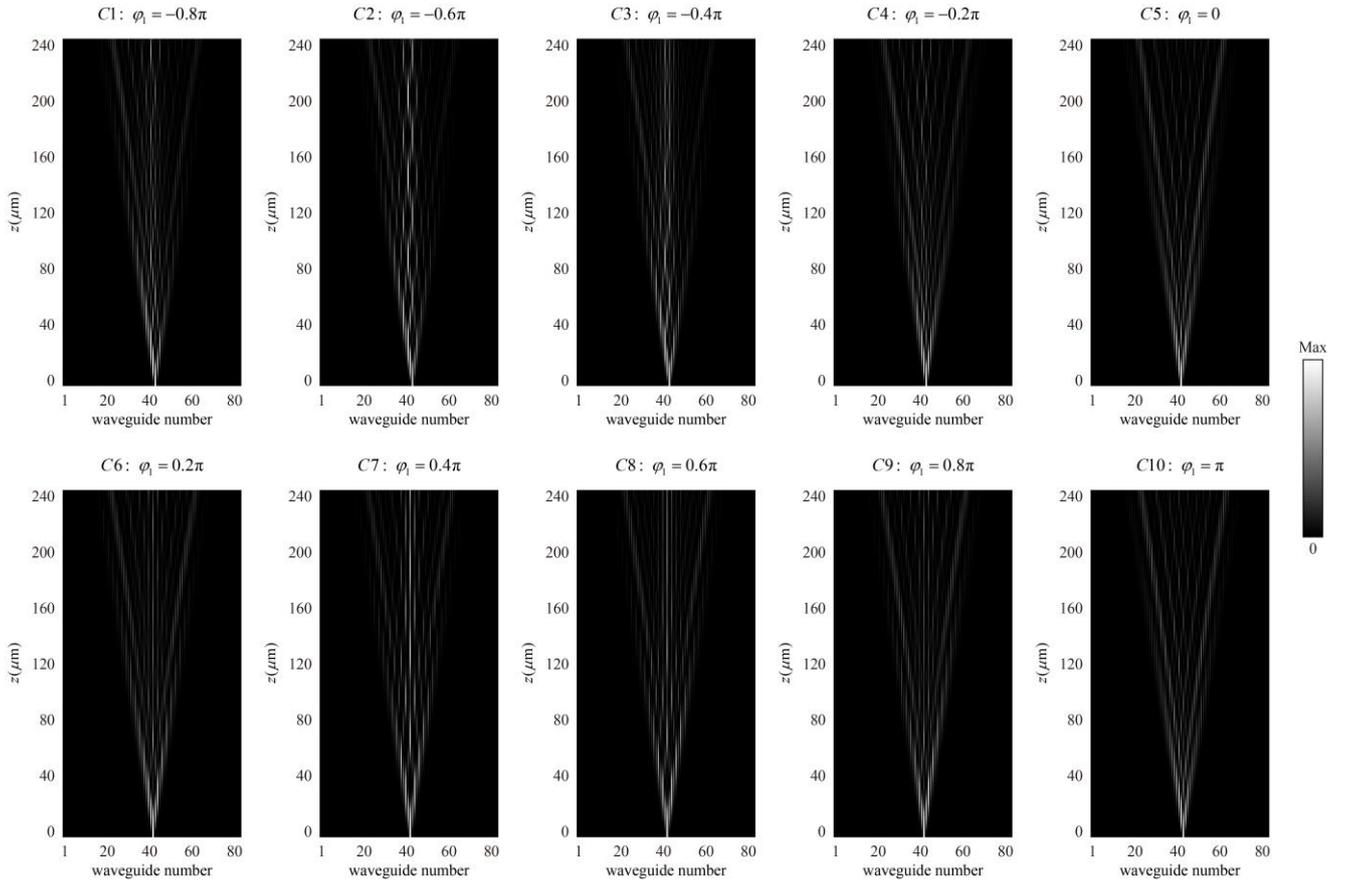

FIG. S11. Propagation simulations for different $\varphi_1$ s in Fig. S10. The interface lies between the waveguide 40 and 41. The gray color range is $[0, 0.2]$.